\documentclass[11pt]{article}
\usepackage{graphicx}
\usepackage{amsmath}
\usepackage{amssymb}
\usepackage{amsxtra}

\usepackage{cite}

\title{Reality from maximizing overlap in the future-included theories}
\author{Keiichi Nagao\\Faculty of Education, Ibaraki University, Bunkyo 2-1-1, \\
Mito, Ibaraki 310-8512 
Japan, \\
email: keiichi.nagao.phys@vc.ibaraki.ac.jp \\
and\\
Holger Bech Nielsen\\Niels Bohr Institute, University of Copenhagen, 
Blegdamsvej 17, \\
Copenhagen $\O$, Denmark, \\
email: hbech@nbi.dk}
\begin{document}
\maketitle

\begin{abstract}

In the future-included complex and real action 
theories whose paths run over not only the past but also the future, 
we briefly review the theorem on the normalized matrix element of an operator $\hat{\cal O}$, 
which is defined in terms of the future and past states 
with a proper inner product $I_Q$ that makes a given Hamiltonian normal. 
The theorem  states that, 
provided that the operator $\hat{\cal O}$ is $Q$-Hermitian, i.e. 
Hermitian with regard to the proper inner product $I_Q$, 
the normalized matrix element 
becomes real and time-develops under 
a $Q$-Hermitian Hamiltonian 
for the past and future states selected such that 
the absolute value of the transition amplitude from 
the past state to the future state is maximized. 
Discussing what the theorem implicates, we speculate that 
the future-included complex action theory 
would be the most elegant quantum theory.

\end{abstract}

\section{Introduction}\label{s:intro}

Quantum theory is formulated via the Feynman path integral (FPI). 
Usually an action in the FPI is taken to be real. 
However, there is a possibility that the action is complex at the fundamental level but looks 
real effectively. 
If we pursue a fundamental theory, it is better to require less conditions imposed on it at first. 
In this sense such a complex action theory (CAT) is preferable to the usual real action theory (RAT), 
because the former has less conditions at least by one: there is no reality condition on the action. 
Based on this speculation the CAT has been investigated 
with the expectation that the imaginary part of the action 
would give some falsifiable predictions~\cite{Bled2006,Nielsen:2007ak,Nielsen:2008cm,Nielsen:2005ub}, 
and various interesting suggestions have been made for Higgs mass~\cite{Nielsen:2007mj}, 
quantum mechanical philosophy~\cite{newer1,Vaxjo2009,newer2}, 
some fine-tuning problems~\cite{Nielsen2010qq,degenerate}, 
black holes~\cite{Nielsen2009hq}, 
de Broglie-Bohm particles and a cut-off in loop diagrams~\cite{Bled2010B}. 
Also, in Ref.~\cite{Nagao:2010xu}, 
introducing what we call the proper inner product $I_Q$ 
so that a given non-normal Hamiltonian becomes normal with respect to it, 
we proposed a mechanism to effectively 
obtain a Hamiltonian which is $Q$-Hermitian, i.e., 
Hermitian with respect to the proper inner product, 
after a long time development. 
Furthermore, using the complex coordinate 
formalism~\cite{Nagao:2011za}, 
we explicitly 
derived the momentum relation $p=m \dot{q}$, where $m$ is a complex mass, 
via the FPI~\cite{Nagao:2011is}. 
In general, the CAT\footnote{The corresponding Hamiltonian $\hat{H}$ is generically non-normal.  So the set of the Hamiltonians we consider is much larger 
than that of the PT-symmetric non-Hermitian Hamiltonians, 
which has been intensively 
studied~\cite{Bender:1998ke,Bender:1998gh,Bender:2011ke,Mostafazadeh_CPT_ip_2002,
Mostafazadeh_CPT_ip_2003}. }  could be classified into two types: 
one is the future-not-included theory~\cite{Nagao:2013eda}\footnote{
In our recent study~\cite{Nagao:2017ecx}, 
we have pointed out that, 
if a theory is described with a complex action,  then such a theory 
is suggested to be the future-included theory 
rather than the future-not-included theory.}, i.e., 
the theory including only a past time as an integration interval of time, 
and the other one is the future-included theory\cite{Bled2006}, 
in which not only the past state $| A(T_A) \rangle$ at the initial time $T_A$ 
but also the future state $| B(T_B) \rangle$ at the final time $T_B$ is given at first, 
and the time integration is performed over the whole period 
from the past to the future.

In the future-included theory, 
the normalized matrix element~\cite{Bled2006} 
$\langle \hat{\cal O} \rangle^{BA} 
\equiv \frac{ \langle B(t) |  \hat{\cal O}  | A(t) \rangle }{ \langle B(t) | A(t) \rangle }$, 
where $t$ is an arbitrary time ($T_A \leq t \leq T_B$), 
seems to have a role of an expectation value of the operator $\hat{\cal O}$. 
Indeed, in Refs.~\cite{Nagao:2012mj,Nagao:2012ye} 
we argued in the case of the action being complex that,  
if we regard $\langle \hat{\cal O} \rangle^{BA}$ 
as an expectation value in the future-included theory, 
we obtain the Heisenberg equation, Ehrenfest's theorem, 
and a conserved probability current density. 
So $\langle \hat{\cal O} \rangle^{BA}$ is a 
strong candidate for the expectation value 
in the future-included theory. 
The normalized matrix element $\langle \hat{\cal O} \rangle^{BA}$ 
is called the weak value~\cite{AAV} in the context of the future-included RAT, 
and it has been intensively studied. 
The details are found in Ref.~\cite{review_wv} and references therein.

In Ref.~\cite{Nagao:2015bya}, 
we considered a slightly modified normalized matrix element 
$\langle \hat{\cal O} \rangle_Q^{BA} 
\equiv \frac{ \langle B(t) |_Q  \hat{\cal O}  | A(t) \rangle }{ \langle B(t) |_Q A(t) \rangle }$, 
where $\langle B(t)|_Q\equiv \langle B(t)|Q$, 
and $Q$ is a Hermitian operator that is appropriately chosen to define the proper inner product $I_Q$. 
This matrix element is obtained just by changing the notation of $\langle B(t)|$ 
as $\langle B(t)| \rightarrow \langle B(t)|_Q$ in $\langle \hat{\cal O} \rangle^{BA}$. 
We proposed a theorem in the future-included CAT, which states that, 
provided that an operator $\hat{\cal O}$ is $Q$-Hermitian, 
$\langle \hat{\cal O} \rangle_Q^{BA}$ becomes real and 
time-develops under a $Q$-Hermitian Hamiltonian for the future and past states 
selected such that the absolute value of the transition amplitude defined with $I_Q$ 
from the past state to the future state is maximized.
We call this way of thinking the maximization principle. 
This theorem applies to not only the CAT but also the RAT. 
In Ref.~\cite{Nagao:2015bya}, we proved this theorem only in the CAT, i.e.,  
in the case of non-Hermitian 
Hamiltonians, by finding that 
essentially only terms associated with the largest imaginary parts 
of the eigenvalues of the Hamiltonian $\hat{H}$\footnote{In the CAT the imaginary parts of the eigenvalues 
of $\hat{H}$ are supposed to be bounded from above 
to avoid the Feynman path integral $\int e^{\frac{i}{\hbar}S} {\cal D} \text{path}$ 
being divergently meaningless. } contribute significantly 
to the absolute value of the transition amplitude defined with $I_Q$, 
and that $\langle \hat{\cal O} \rangle_Q^{BA}$
for such maximizing states becomes an expression similar to 
an expectation value defined with $I_Q$ in the future-not-included theory. 
This proof is based on the existence of imaginary parts of the 
eigenvalues of $\hat{H}$, 
so it cannot be applied to the RAT. 
In Ref.~\cite{Nagao:2017cpl}, we presented another theorem 
particular to the case of Hermitian Hamiltonians, i.e., the RAT case for simplicity, 
and proved it. 
In this paper, we review the maximization principle 
and clarify what the theorems implicate 
based on Refs.~\cite{Nagao:2015bya, Nagao:2017cpl, Nagao:2017book}.

This paper is organized as follows. 
In section 2 we briefly review the proper inner product and the future-included theory. 
In section 3 we present the theorems, and prove them in section 4. 
Section 5 is devoted to discussion.


\section{Proper inner product and future-included complex action theory}

We suppose that our system that 
could be the whole world is described by a non-normal diagonalizable 
Hamiltonian $\hat{H}$ such that $[\hat{H}, \hat{H}^{\dagger}]\neq 0$. 
Based on Refs.\cite{Nagao:2010xu,Nagao:2011za,Nagao:2017book}, we first
review the proper inner product for $\hat{H}$ which makes $\hat{H}$ 
normal with respect to it. 
We define the eigenstates
 $| \lambda_i \rangle (i=1,2,\cdots)$ of $\hat{H}$ such that 
\begin{equation}
\hat{H} | \lambda_i \rangle = \lambda_i | \lambda_i \rangle, 
\end{equation}
where $\lambda_i (i=1,2,\cdots)$ are the eigenvalues of 
$\hat{H}$, and introduce the diagonalizing operator 
$P=(| \lambda_1 \rangle , | \lambda_2 \rangle , \ldots)$, 
so that $\hat{H}$ is diagonalized as 
$\hat{H} = PD P^{-1}$, 
where $D$ is given by $\text{diag}(\lambda_1, \lambda_2, \cdots)$. 
Let us consider a transition from an eigenstate 
$| \lambda_i \rangle$ to another 
$| \lambda_j \rangle ~(i \neq j)$ fast in time $\Delta t$.  
Since $| \lambda_i \rangle$ are not orthogonal to 
each other in the usual inner product $I$, 
$I(| \lambda_i \rangle , | \lambda_j \rangle ) \equiv \langle \lambda_i | \lambda_j \rangle \neq \delta_{ij}$, 
the transition can be measured, i.e., 
$|I(| \lambda_j \rangle, \exp\left( -\frac{i}{\hbar} \hat{H} \Delta t \right) |\lambda_i \rangle )|^2 \neq 0$, 
though $\hat{H}$ cannot bring the system from $| \lambda_i \rangle$ 
to $| \lambda_j \rangle ~(i \neq j)$. 
In any reasonable theories, such an unphysical transition from an eigenstate to another one 
with a different eigenvalue should be prohibited. 
In order to have reasonable probabilistic results, 
we introduce 
a proper inner product \cite{Nagao:2010xu, Nagao:2011za}\footnote{Similar inner products are studied also 
in refs.\cite{Geyer,Mostafazadeh_CPT_ip_2002,Mostafazadeh_CPT_ip_2003}.} 
for arbitrary kets $|u \rangle$ and $|v \rangle$ as 
\begin{equation}
I_Q(|u \rangle , |v \rangle) \equiv \langle u |_Q v \rangle 
\equiv \langle u | Q | v \rangle,
\end{equation}  
where $Q$ is a Hermitian operator chosen as 
$Q=(P^\dag)^{-1} P^{-1}$, 
so that $| \lambda_i \rangle$ get orthogonal to each other 
with regard to $I_Q$, 
\begin{equation} 
\langle \lambda_i  |_Q \lambda_j \rangle = \delta_{ij}. 
\end{equation}
This implies the orthogonality relation 
$\sum_i | \lambda_i \rangle \langle \lambda_i |_{Q} = 1$. 
In the special case of $\hat{H}$ being hermitian, $Q$ is the unit operator. 
We introduce the ``$Q$-Hermitian'' conjugate $\dag^Q$ of 
an operator $A$ by 
$\langle u |_Q A | v \rangle^* \equiv \langle v |_Q A^{\dag^Q} | u \rangle$, so 
\begin{equation} 
A^{\dag^Q} \equiv Q^{-1} A^\dag Q. 
\end{equation}
If $A$ obeys $A^{\dag^Q} = A$, $A$ is $Q$-Hermitian. 
We also define $\dag^Q$ for kets and bras as 
$| u \rangle^{\dag^Q} \equiv \langle u |_Q $ and 
$\left(\langle u |_Q \right)^{\dag^Q} \equiv | u \rangle$. 
In addition, 
$P^{-1}=
\left(
 \begin{array}{c}
      \langle \lambda_1 |_Q     \\
      \langle \lambda_2 |_Q     \\
      \vdots 
 \end{array}
\right)$ 
satisfies $P^{-1} \hat{H} P = D$ and 
$P^{-1} \hat{H}^{\dag^Q} P = D^{\dag}$, 
so $\hat{H}$ is ``$Q$-normal'', 
$[\hat{H}, \hat{H}^{\dag^Q} ] = P [D, D^\dag ] P^{-1} =0$.  
Thus the inner product $I_Q$ makes $\hat{H}$ 
$Q$-normal. 
We note that $\hat{H}$ can be decomposed as 
$\hat{H}=\hat{H}_{Qh} + \hat{H}_{Qa}$, 
where $\hat{H}_{Qh}= \frac{\hat{H} + \hat{H}^{\dag^Q} }{2}$ and 
$\hat{H}_{Qa} = \frac{\hat{H} - \hat{H}^{\dag^Q} }{2}$ are 
$Q$-Hermitian and anti-$Q$-Hermitian parts of $\hat{H}$ respectively.

In Refs.\cite{Bled2006,Nagao:2012mj,Nagao:2012ye}, 
the future-included theory is 
described by using 
the future state $| B (T_B) \rangle$ at the final time $T_B$ 
and the past state $| A (T_A) \rangle$ at the initial time $T_A$, 
where $| A (T_A) \rangle$ and $| B (T_B) \rangle$ 
time-develop as follows, 
\begin{eqnarray}
&&i \hbar \frac{d}{dt} | A(t) \rangle = \hat{H} | A(t) \rangle , \label{schro_eq_Astate} \\
&&-i \hbar \frac{d}{dt} \langle B(t) |  
= \langle B(t) |  \hat{H} , \label{schro_eq_Bstate_old} 
\end{eqnarray}
and the normalized matrix element 
$\langle \hat{\cal O} \rangle^{BA} 
\equiv \frac{ \langle B(t) |  \hat{\cal O}  | A(t) \rangle }{ \langle B(t) | A(t) \rangle }$ 
is studied. 
The quantity $\langle \hat{\cal O} \rangle^{BA}$ 
is called the weak value\cite{AAV, review_wv} in the RAT.  
In refs.\cite{Nagao:2012mj,Nagao:2012ye}, 
we investigated $\langle \hat{\cal O} \rangle^{BA}$ and found that, 
if we regard $\langle \hat{\cal O} \rangle^{BA}$ 
as an expectation value in the future-included theory, 
then we obtain the Heisenberg equation, Ehrenfest's theorem, 
and a conserved probability current density. 
Therefore, $\langle \hat{\cal O} \rangle^{BA}$ seems to have a 
role of an expectation value in the future-included theory.  

In the following, we adopt the proper inner product $I_Q$ 
for all quantities. Hence we change the notation of  the final state $\langle B(T_B) |$ 
as $\langle B(T_B) | \rightarrow \langle B(T_B) |_Q$ 
so that the Hermitian operator $Q$ pops out 
and the usual inner product $I$ is replaced with $I_Q$. 
Then $\langle B(T_B) |$ time-develops according not to 
eq.(\ref{schro_eq_Bstate_old}) but to 
\begin{eqnarray}
-i \hbar \frac{d}{dt} \langle B(t) |_Q  
= \langle B(t) |_Q  \hat{H} 
\quad \Leftrightarrow \quad 
i \hbar \frac{d}{dt} | B(t) \rangle = {\hat{H}}^{\dag^Q} | B(t) \rangle ,  
\label{schro_eq_Bstate} 
\end{eqnarray}
and the normalized matrix element is expressed as 
\begin{equation}
\langle \hat{\cal O} \rangle_Q^{BA} 
\equiv \frac{ \langle B(t) |_Q  \hat{\cal O}  | A(t) \rangle }{ \langle B(t) |_Q A(t) \rangle }.  
\end{equation} 
In addition, we suppose that $|A(T_A) \rangle$ and $\langle B(T_B)|$ are $Q$-normalized by 
$\langle A(T_A) |_{Q} A(T_A) \rangle =1$ and $\langle B(T_B) |_{Q} B(T_B) \rangle = 1$. 
In the RAT, since $Q=1$, $\langle \hat{\cal O} \rangle_Q^{BA}$ corresponds to 
$\langle \hat{\cal O} \rangle^{BA}$.


\section{Theorems of the maximization principle}

In Ref.~\cite{Nagao:2015bya} we proposed the following theorem :

\vspace{0.5cm}

\noindent
{\bf Theorem 1. Maximization principle in the future-included CAT} \\ 
{\em 
As a prerequisite, assume that a given Hamiltonian 
$\hat{H}$ is non-normal but diagonalizable 
and that the imaginary parts of the eigenvalues 
of $\hat{H}$ are bounded from above, 
and define a modified inner product $I_Q$ by means 
of a Hermitian operator $Q$ arranged so 
that $\hat{H}$ becomes normal with respect to $I_Q$. 
Let the two states $| A(t) \rangle$ and $ | B(t) \rangle$ 
time-develop according to the Schr\"{o}dinger equations 
with $\hat{H}$ and $\hat{H}^{\dag^Q}$ respectively: 
$|A (t) \rangle = e^{-\frac{i}{\hbar}\hat{H} (t-T_A) }| A(T_A) \rangle$, 
$|B (t) \rangle = e^{-\frac{i}{\hbar} {\hat{H}}^{\dag^Q} (t-T_B) } | B(T_B)\rangle$, 
and be normalized with $I_Q$ 
at the initial time $T_A$ and the final time $T_B$ respectively: 
$\langle A(T_A) |_{Q} A(T_A) \rangle = 1$,
$\langle B(T_B) |_{Q} B(T_B) \rangle = 1$. 
Next determine $|A(T_A) \rangle$ and $|B(T_B) \rangle$ so as to maximize 
the absolute value of the transition amplitude 
$|\langle B(t) |_Q A(t) \rangle|=|\langle B(T_B)|_Q \exp(-i\hat{H}(T_B-T_A)) |A(T_A) \rangle|$. 
Then, provided that an operator $\hat{\cal O}$ is $Q$-Hermitian, i.e., Hermitian 
with respect to the inner product $I_Q$, 
$\hat{\cal O}^{\dag^Q} = \hat{\cal O}$, 
the normalized matrix element of the operator $\hat{\cal O}$ defined by 
$\langle \hat{\cal O} \rangle_Q^{BA} 
\equiv
\frac{\langle B(t) |_Q \hat{\cal O} | A(t) \rangle}{\langle B(t) |_Q A(t) \rangle}$ 
becomes {\rm real} and time-develops under 
a {\rm $Q$-Hermitian} Hamiltonian. }


\vspace*{0.5cm}


We call this way of thinking the maximization principle. 
This theorem means that the normalized matrix element 
$\langle \hat{\cal O} \rangle_Q^{BA}$, which is taken as an average 
for an operator $\hat{\cal O}$ obeying $\hat{\cal O}^{\dag^Q} =\hat{\cal O}$, 
turns out to be real almost unavoidably. 
Also, in the case of non-normal Hamiltonians, it is nontrivial to obtain the 
emerging $Q$-hermiticity for the Hamiltonian by the maximization principle. 
The theorem is given for systems defined with such general Hamiltonians 
that they do not even have to be normal, so it can 
also be used for normal Hamiltonians 
in addition to non-normal Hamiltonians. 
For a normal Hamiltonian $\hat{H}$, $Q$ is the unit operator. 
In such a case the above theorem becomes simpler with $Q=1$. 
There are two possibilities for such a case: 
one is that $\hat{H}$ is non-Hermitian but normal, and the other is that $\hat{H}$ is 
Hermitian. 
In both cases $Q=1$, but there is a significant difference between them. 
In the former case, there are imaginary parts of the eigenvalues of $\hat{H}$, 
$\text{Im}\lambda_i$, and the eigenstates having the largest 
$\text{Im}\lambda_i$ blow up and contribute most to the the absolute value 
of the transition amplitude $|\langle B(t) |_Q A(t) \rangle|$. 
In the latter case, there are no $\text{Im}\lambda_i$, 
and the full set of the eigenstates of $\hat{H}$ can contribute to 
$| \langle B (t) | A (t) \rangle |$. 
So we need to investigate them separately.

In the special case where the Hamiltonian is Hermitian, i.e., in the future-included RAT, 
we can consider three possibilities: 
One is that  $ | A(T_A) \rangle$ is given at first, and $ | B(T_B) \rangle$ is chosen 
by the maximization principle. Another is the reverse. The other is that both 
$| A(T_A) \rangle$ and $ | B(T_B) \rangle$ are partly given and chosen. 
Since we know empirically the second law of thermodynamics, 
we choose the first option in the future-included RAT. 
We suppose that $| A(t) \rangle$ is a given 
fixed state, and only $| B(t) \rangle$ is a random state, which should be chosen appropriately 
by the maximization principle, though in the future-included CAT 
both $| A(t) \rangle$ and $| B(t) \rangle$ are supposed to be random states at first. 
In addition,  in the future-included RAT the hermiticity of the Hamiltonian is given at first, 
so we write the theorem particular to 
the case of Hermitian Hamiltonians as follows:

\vspace{0.5cm}

\noindent
{\bf Theorem 2. Maximization principle in the future-included RAT} \\ 
{\em 
As a prerequisite, assume that a given Hamiltonian 
$\hat{H}$ is diagonalizable and Hermitian.  
Let the two states $| A(t) \rangle$ and $ | B(t) \rangle$ 
time-develop according to the Schr\"{o}dinger equation with $\hat{H}$: 
$|A (t) \rangle = e^{-\frac{i}{\hbar}\hat{H} (t-T_A) }| A(T_A) \rangle$, 
$|B (t) \rangle = e^{-\frac{i}{\hbar} {\hat{H}} (t-T_B) }$ $| B(T_B)\rangle$, 
and be normalized at the initial time $T_A$ and the final time $T_B$ respectively: 
$\langle A(T_A) | A(T_A) \rangle = 1$, 
$\langle B(T_B) | B(T_B) \rangle = 1$. 
Next determine $|B(T_B) \rangle$ for the given $|A(T_A) \rangle$ so as to maximize 
the absolute value of the transition amplitude 
$|\langle B(t) | A(t) \rangle|=
|\langle B(T_B)| \exp(-\frac{i}{\hbar}\hat{H}(T_B-T_A)) | A(T_A) \rangle|$. 
Then, provided that an operator $\hat{\cal O}$
is Hermitian, $\hat{\cal O}^\dag = \hat{\cal O}$, 
the normalized matrix element of the operator $\hat{\cal O}$ defined by 
$\langle \hat{\cal O} \rangle^{BA} 
\equiv
\frac{\langle B(t) | \hat{\cal O} | A(t) \rangle}{\langle B(t) | A(t) \rangle}$ 
becomes {\rm real} and time-develops under the given Hermitian Hamiltonian. }

\vspace{0.5cm}

We investigate the above theorems separately.

\section{Proof of the theorems}

To prove the theorems 
we expand $| A(t) \rangle$ and $| B(t) \rangle$ in terms of the eigenstates $| \lambda_i \rangle$ 
as follows: 
\begin{eqnarray}
&&|A (t) \rangle = \sum_i a_i (t) | \lambda_i \rangle, \label{Aketexpansion}\\
&&|B (t) \rangle = \sum_i b_i (t) | \lambda_i \rangle,  \label{Bketexpansion}
\end{eqnarray}
where 
\begin{eqnarray}
&&a_i (t) = a_i (T_A) e^{-\frac{i}{\hbar}\lambda_i (t-T_A) }, \label{aitimedevelopment} \\ 
&&b_i (t) = b_i (T_B) e^{-\frac{i}{\hbar}\lambda_i^* (t-T_B) }. \label{bitimedevelopment} 
\end{eqnarray}
We express  $a_i(T_A)$ and $b_i(T_B)$ as 
\begin{eqnarray}
&&a_i(T_A)= | a_i(T_A) | e^{i \theta_{a_i}}, \label{aiTA} \\
&&b_i(T_B) = | b_i(T_B) | e^{i \theta_{b_i}}, \label{biTB}
\end{eqnarray}
and introduce 
\begin{eqnarray}
&&T\equiv T_B - T_A , \label{T} \\
&&\Theta_i \equiv \theta_{a_i} - \theta_{b_i} - \frac{1}{\hbar} T \text{Re} \lambda_i , \label{Thetai} \\
&&R_i \equiv |a_i (T_A)| |b_i (T_B)| e^{\frac{1}{\hbar} T \text{Im} \lambda_i }. \label{RiImLi}
\end{eqnarray} 
Then, since $\langle B (t) |_Q A (t) \rangle$ is expressed as 
\begin{equation}
\langle B (t) |_Q A (t) \rangle = \sum_i R_i e^{i \Theta_i}, \label{BbraQAket}
\end{equation}
$| \langle B (t) |_Q A (t) \rangle |^2$ is calculated as 
\begin{equation}
| \langle B (t) |_Q A (t) \rangle |^2
= \sum_i R_i^2 + 2 \sum_{i<j} R_i R_j \cos(\Theta_i - \Theta_j). \label{|BbraQAket|^2}
\end{equation}
The normalization conditions for $| A(T_A) \rangle$ and $| B(T_B) \rangle$ 
are expressed as 
\begin{equation}
\sum_i  | a_i (T_A) |^2  = \sum_i  | b_i (T_B) |^2  = 1. \label{nc_ATABTB} 
\end{equation}  

We proceed with this study separately according to whether the given Hamiltonian $\hat{H}$ is  
non-Hermitian or Hermitian.

\subsection{Non-Hermitian Hamiltonians case}\label{subsec:non-Hermitian}

In the case of non-Hermitian Hamiltonians, there exist imaginary parts of the 
eigenvalues of the Hamiltonian, $\text{Im} \lambda_i$, 
which are supposed to be bounded from above 
to avoid the Feynman path integral $\int e^{\frac{i}{\hbar}S} {\cal D} \text{path}$ 
being divergently meaningless. 
We can imagine that some of $\text{Im} \lambda_i$ 
take the maximal value $B$, and denote 
the corresponding subset of $\{ i \}$ as $A$. 
Then, since $R_i \geq 0$, $| \langle B (t) |_Q A (t) \rangle |$ can take 
a maximal value only under the following conditions: 
\begin{eqnarray}
&& | a_i (T_A) |  = | b_i (T_B) | =0 \quad \text{for $\forall i \notin A$} , \label{abinotinA0} \\
&& \Theta_i  
\equiv \Theta_c 
\quad \text{for $\forall i \in A$} \label{max_cond_theta} , \\ 
&& \sum_{i \in A} | a_i (T_A) |^2 =\sum_{i \in A}|b_i (T_B)|^2 = 1,  \label{nc_ATABTB3} 
\end{eqnarray}
and $| \langle B (t) |_Q A (t) \rangle |^2$ is estimated as 
\begin{eqnarray}
| \langle B (t) |_Q A (t) \rangle |^2
&=& \left( \sum_{i \in A} R_i \right)^2  \nonumber \\  
&=& e^{\frac{2 B T}{\hbar} } 
\left( \sum_{i \in A} |a_i (T_A)| |b_i (T_B)| \right)^2   \nonumber \\
&\leq& e^{\frac{2 B T}{\hbar} } 
\left\{ \sum_{i \in A} \left( \frac{ |a_i (T_A)| + |b_i (T_B)|}{2} \right)^2 \right\}^2  \nonumber \\
&=& e^{\frac{2}{\hbar} B T} , 
\end{eqnarray}
where the third equality is realized for 
\begin{equation}
 |a_i (T_A)| = |b_i (T_B)|  \quad \text{for $\forall i \in A$}. 
\label{max_cond_ab}  \
\end{equation}
In the last equality we have used this relation 
and Eq.(\ref{nc_ATABTB3}). 
The maximization condition of  $| \langle B (t) |_Q A (t) \rangle |$ 
is represented by 
Eqs.(\ref{abinotinA0})-(\ref{nc_ATABTB3}) and (\ref{max_cond_ab}). 
That is to say, 
the states to maximize $| \langle B (t) |_Q A (t) \rangle |$, 
$| A(t) \rangle_{\rm{max}}$ and $| B(t) \rangle_{\rm{max}}$, are expressed as 
\begin{eqnarray}
&&|A (t) \rangle_{\rm{max}} = \sum_{i \in A} a_i (t) | \lambda_i \rangle , 
\label{Atketmax_sum_inA_ai} \\
&&|B (t) \rangle_{\rm{max}} = \sum_{i \in A} b_i (t) | \lambda_i \rangle, 
\label{Btketmax_sum_inA_bi} 
\end{eqnarray}
where $a_i (t)$ and $b_i (t)$ obey 
Eqs.(\ref{max_cond_theta}), (\ref{nc_ATABTB3}), and (\ref{max_cond_ab}).

To evaluate $\langle \hat{\cal O} \rangle_Q^{BA}$ 
for $| A(t) \rangle_{\rm{max}} $ and $| B(t) \rangle_{\rm{max}}$,  
utilizing the $Q$-Hermitian part of $\hat{H}$, $\hat{H}_{Qh} \equiv \frac{\hat{H} + \hat{H}^{\dag^Q} }{2}$, 
we define the following state: 
\begin{equation}
| \tilde{A}(t) \rangle \equiv 
e^{-\frac{i}{\hbar}(t-T_A) \hat{H}_{Qh}} | A(T_A) \rangle_{\rm{max}}, 
\end{equation}
which is normalized as 
$\langle \tilde{A}(t) |_Q \tilde{A}(t) \rangle = 1$ 
and obeys the Schr\"{o}dinger equation 
\begin{eqnarray}
i\hbar  \frac{d}{d t}| \tilde{A}(t) \rangle 
&=& \hat{H}_{Qh} | \tilde{A}(t) \rangle .  \label{ScheqAtildetket}
\end{eqnarray} 
Using Eqs.(\ref{abinotinA0})-(\ref{nc_ATABTB3}) and (\ref{max_cond_ab}), we obtain 
\begin{equation} 
{}_{\rm{max}} \langle B (t) |_Q A (t) \rangle_{\rm{max}} 
=
e^{i \Theta_c} \sum_{i \in A} R_i  
= e^{i \Theta_c} e^{\frac{B T}{\hbar} } , 
\end{equation}
and 
\begin{eqnarray} 
&&{}_{\rm{max}} \langle B (t) |_Q \hat{\cal O} | A  (t) \rangle_{\rm{max}} \nonumber \\
&=& 
e^{i \Theta_c} e^{\frac{B T}{\hbar} }
\sum_{i , j \in A} a_j(T_A)^*  a_i(T_A) 
e^{\frac{i}{\hbar}(t-T_A) (\text{Re}\lambda_j - \text{Re}\lambda_i)} 
\langle \lambda_j |_Q \hat{\cal O} | \lambda_i \rangle  \nonumber \\
&=& 
e^{i \Theta_c} e^{\frac{B T}{\hbar} }
\langle \tilde{A}(t) |_Q \hat{\cal O} | \tilde{A}(t) \rangle .
\end{eqnarray}
Thus 
$\langle \hat{\cal O} \rangle_Q^{BA}$  
for $| A(t) \rangle_{\rm{max}}$ and $| B(t) \rangle_{\rm{max}}$ 
is evaluated as 
\begin{eqnarray}
\langle \hat{\cal O} \rangle_Q^{B_{\rm{max}} A_{\rm{max}}} 
&=& 
\langle \tilde{A}(t) |_Q \hat{\cal O} | \tilde{A}(t) \rangle 
\equiv 
\langle \hat{\cal O} \rangle_Q^{\tilde{A} \tilde{A}} . \label{OBAmaxtilde}
\end{eqnarray}
Since 
$\left\{ \langle \hat{\cal O} \rangle_Q^{\tilde{A} \tilde{A}} \right\}^*=\langle \hat{\cal O}^{\dag^Q} \rangle_Q^{\tilde{A} \tilde{A}}$, 
$\langle \hat{\cal O} \rangle_Q^{BA}$ 
for $| A(t) \rangle_{\rm{max}} $ and $| B(t) \rangle_{\rm{max}}$ 
has been shown to be real for $Q$-Hermitian $\hat{\cal O}$.

Next we study the time development of 
$\langle \hat{\cal O} \rangle_Q^{\tilde{A} \tilde{A}}$. 
We express $\langle \hat{\cal O} \rangle_Q^{\tilde{A} \tilde{A}}$ as 
$\langle \hat{\cal O} \rangle_Q^{\tilde{A} \tilde{A}}
=\langle \tilde{A}(T_A) |_Q \hat{\cal O}_{H}(t, T_A)
 | \tilde{A}(T_A) \rangle$, 
where we have introduced the Heisenberg operator 
$\hat{\cal O}_{H}(t, T_A) 
\equiv 
e^{ \frac{i}{\hbar} \hat{H}_{Qh} (t-T_A) } 
\hat{\cal O} 
e^{ -\frac{i}{\hbar} \hat{H}_{Qh} (t-T_A)}$. 
This operator $\hat{\cal O}_{H}(t, T_A)$ obeys 
the Heisenberg equation 
$i\hbar  \frac{d}{d t} \hat{\cal O}_{H}(t, T_A) 
= [ \hat{\cal O}_{H}(t, T_A) , \hat{H}_{Qh} ]$, 
so we find that 
$\langle \hat{\cal O} \rangle_Q^{\tilde{A} \tilde{A}}$ time-develops 
under the $Q$-Hermitian Hamiltonian $\hat{H}_{Qh}$ as 
\begin{eqnarray}
\frac{d}{dt} \langle \hat{\cal O} \rangle_Q^{\tilde{A} \tilde{A}} 
&=&
\frac{i}{\hbar} 
\langle \left[ \hat{H}_{Qh}, \hat{\cal O} \right]  
\rangle_Q^{\tilde{A} \tilde{A}} . 
\label{ddtOAtildeAtildeQ}
\end{eqnarray}

Thus Theorem 1 has been proven, and the maximization principle provides both 
the reality of $\langle \hat{\cal O} \rangle_Q^{BA}$ 
for $Q$-Hermitian $\hat{\cal O}$ and the $Q$-Hermitian Hamiltonian.

\subsection{Hermitian Hamiltonians case}\label{subsec:Hermitian}

Theorem 2 can be proven more simply than Theorem 1. 
Since the norms of $ |  A(t) \rangle$ and $ |  B(t) \rangle$ are constant in time 
in the case of Hermitian Hamiltonians,  
\begin{eqnarray}
&&\langle A(t) |  A(t) \rangle=\langle A(T_A) |  A(T_A) \rangle=1,  \label{normAconstant} \\
&&\langle B(t) |  B(t) \rangle=\langle B(T_B) |  B(T_B) \rangle=1, 
\end{eqnarray}
we can directly use an elementary property of linear space, 
and find that the final state to maximize $| \langle B (t) | A (t) \rangle |$, 
$| B(T_B) \rangle_{\rm{max}}$, 
is the same as $| A(t) \rangle$ up to a constant phase factor:   
\begin{eqnarray}
|B (t) \rangle_{\rm{max}} 
= 
e^{-i \Theta_c } | A(t) \rangle . \label{BmaxphaseA}
\end{eqnarray}
This phase factor presents the ambiguity of the maximizing state  $| B(t) \rangle_{\rm{max}}$, 
and shows that $| B(t) \rangle_{\rm{max}}$ is not determined uniquely. 
We note that this is quite in contrast to the case of non-Hermitian Hamiltonians, where 
only a unique class of $| A(t) \rangle$ and $| B(t) \rangle$ is chosen by the maximization principle.  
The normalized matrix element $\langle \hat{\cal O} \rangle^{BA}$ 
for the given $| A(t) \rangle$ and $| B(t) \rangle_{\rm{max}}$ becomes 
\begin{eqnarray}
\langle \hat{\cal O} \rangle^{B_{\rm{max}}A} 
&=&
\frac{ {}_{\rm{max}}\langle B(t) |  \hat{\cal O}  | A(t) \rangle }{ {}_{\rm{max}}\langle B(t) | A(t) \rangle } \nonumber \\
&=&
\langle A(t) |  \hat{\cal O}  | A(t) \rangle \nonumber \\
&\equiv&
\langle\hat{\cal O} \rangle^{AA} , 
\end{eqnarray} 
where in the second equality we have used Eqs.(\ref{BmaxphaseA}) and (\ref{normAconstant}). 
Thus $\langle \hat{\cal O} \rangle^{BA}$ 
for the given $| A(t) \rangle$ and $| B(t) \rangle_{\rm{max}}$ 
has become the form of a usual average $\langle \hat{\cal O} \rangle^{AA}$, 
and so it becomes real for Hermitian $\hat{\cal O}$. 
In addition, $\langle \hat{\cal O} \rangle^{AA}$ time-develops 
under the Hermitian Hamiltonian $\hat{H}$ as 
\begin{eqnarray}
\frac{d}{dt} \langle \hat{\cal O} \rangle^{A A} 
&=&
\frac{i}{\hbar} 
\langle \left[ \hat{H}, \hat{\cal O} \right]  
\rangle^{A A} . 
\end{eqnarray}
We emphasize that the maximization principle 
provides the reality of $\langle \hat{\cal O} \rangle^{BA}$ for Hermitian $\hat{\cal O}$, 
though $\langle \hat{\cal O} \rangle^{BA}$ is generically complex by definition.

To see the differences from the case of non-Hermitian Hamiltonians more explicitly, 
we investigate Theorem 2 by expanding $| A(t) \rangle$ and $| B(t) \rangle$ 
in the same way as Eqs.(\ref{Aketexpansion})-(\ref{bitimedevelopment}). 
Then we can make use of Eqs.(\ref{aiTA})-(\ref{nc_ATABTB}) 
just by noting that Eqs.(\ref{RiImLi})-(\ref{|BbraQAket|^2}) are expressed as 
\begin{eqnarray}
&&R_i \equiv |a_i (T_A)| |b_i (T_B)| , \\
&&\langle B (t) | A (t) \rangle = \sum_i R_i e^{i \Theta_i} , \\
&& | \langle B (t) | A (t) \rangle |^2
= \sum_i R_i^2 + 2 \sum_{i<j} R_i R_j \cos(\Theta_i - \Theta_j) , 
\end{eqnarray}
since $\text{Im}\lambda_i=0$ and $Q=1$. 
Then, since $R_i \geq 0$, $| \langle B (t) | A (t) \rangle |$ can take 
a maximal value only under the condition: 
\begin{equation}
\Theta_i  
= \Theta_c 
\quad \text{for $\forall i $} \label{max_cond_thetareal} ,  
\end{equation}
and $| \langle B (t) | A (t) \rangle |^2$ is estimated as 
\begin{eqnarray}
| \langle B (t) | A (t) \rangle |^2
&=& \left( \sum_{i } R_i \right)^2  \nonumber \\  
&=& \left( \sum_{i } |a_i (T_A)| |b_i (T_B)| \right)^2   \nonumber \\
&\leq&  
\left\{ \sum_{i} \left( \frac{ |a_i (T_A)| + |b_i (T_B)|}{2} \right)^2 \right\}^2 \nonumber \\ 
&=&1 , 
\end{eqnarray}
where the third equality is realized for 
\begin{equation}
 |a_i (T_A)| = |b_i (T_B)|  \quad \text{for $\forall i $}. 
\label{max_cond_abreal}  
\end{equation}
In the last equality we have used this relation and Eq.(\ref{nc_ATABTB}). 
The condition for maximizing $| \langle B (t) | A (t) \rangle |$ 
is represented by Eqs.(\ref{max_cond_thetareal}) and (\ref{max_cond_abreal}). 
In the case of non-Hermitian Hamiltonians, 
the condition for maximizing $| \langle B (t) |_Q A (t) \rangle |$ 
is represented by 
Eqs.(\ref{abinotinA0})-(\ref{nc_ATABTB3}) and (\ref{max_cond_ab}),  
and essentially only the subset having the 
largest imaginary parts of the eigenvalues of $\hat{H}$ contributes most to 
the absolute value of the transition amplitude $| \langle B (t) |_Q A (t) \rangle |$, 
as we saw in Subsection~\ref{subsec:non-Hermitian}. 
This is quite in contrast to the present study in the case of Hermitian Hamiltonians, 
where the full set of the eigenstates of $\hat{H}$ can contribute to 
$| \langle B (t) | A (t) \rangle |$. 
Thus the final state to maximize $| \langle B (t) | A (t) \rangle |$, 
$| B(T_B) \rangle_{\rm{max}}$, is expressed as   
\begin{eqnarray}
|B (T_B) \rangle_{\rm{max}} 
&=& \sum_{i } b_i^{\rm{max}} (T_B) | \lambda_i \rangle, 
\label{Btketmax_TB} 
\end{eqnarray} 
where 
\begin{eqnarray}
b_i^{\rm{max}} (T_B) 
&\equiv& |a_i (T_A)|  e^{ i \left(  \theta_{a_i} - \frac{1}{\hbar} T \lambda_i - \Theta_c \right) }  
\label{bimaxTB} 
\end{eqnarray}
obeys 
\begin{equation}
\sum_{i }|b_i^{\rm{max}} (T_B)|^2 = 1. 
\end{equation} 
Hence $| B(t) \rangle_{\rm{max}}$ is expressed as 
\begin{eqnarray} 
|B (t) \rangle_{\rm{max}} 
&=& 
e^{-\frac{i}{\hbar} \hat{H}(t-T_B)} | B(T_B) \rangle_{\rm{max}} \nonumber \\ 
&=&\sum_{i } b_i^{\rm{max}} (t) | \lambda_i \rangle, 
\end{eqnarray}
where $b_i^{\rm{max}} (t)$ is given by 
\begin{eqnarray}
b_i^{\rm{max}} (t) 
&=& b_i^{\rm{max}} (T_B) e^{-\frac{i}{\hbar}\lambda_i (t-T_B) } \nonumber \\ 
&=& a_i (t) e^{-i \Theta_c } .  
\end{eqnarray}
In the second equality we have used Eq.(\ref{bimaxTB}). 
Consequently, $| B(t) \rangle_{\rm{max}}$ is found to be the same as $| A(t) \rangle$ 
up to the constant phase factor, as we saw in Eq.(\ref{BmaxphaseA}).

\section{Discussion}

In this paper, after briefly explaining the proper inner product $I_Q$, which makes a given non-normal 
Hamiltonian normal, and also the future-included CAT, 
we have reviewed the theorem on the normalized matrix element of $\hat{\cal O}$, 
$\langle \hat{\cal O} \rangle_Q^{BA}$, which seems to have a 
role of an expectation value in the future-included CAT and RAT. 
Assuming that a given Hamiltonian $\hat{H}$ is 
non-normal but diagonalizable, and that  
the imaginary parts of the eigenvalues of $\hat{H}$ 
are bounded from above, 
we presented a theorem 
that states that, provided that $\hat{\cal O}$ is $Q$-Hermitian, i.e., 
$\hat{\cal O}^{\dag^Q}=\hat{\cal O}$, 
and that $|A(t) \rangle $ and $|B(t) \rangle$ time-develop 
according to the Schr\"{o}dinger equations 
with $\hat{H}$ and $\hat{H}^{\dag^Q}$ and are 
$Q$-normalized 
at the initial time $T_A$ and at the final time $T_B$, respectively, 
$\langle \hat{\cal O} \rangle_Q^{BA}$ becomes real 
and time-develops under a $Q$-Hermitian Hamiltonian  
for $|A(t) \rangle $ and $|B(t) \rangle$ such that the 
absolute value of the transition amplitude 
$|\langle B(t)|_Q A(t) \rangle|$ is maximized. 
First we proved the theorem in the case of non-Hermitian Hamiltonians based on Refs.~\cite{Nagao:2015bya, Nagao:2017book}. 
Next we provided another theorem particular to the case of Hermitian Hamiltonians, and proved it,  
based on Refs.~\cite{Nagao:2017cpl, Nagao:2017book}.  
It is noteworthy that, both in the future-included CAT and RAT, 
we have obtained a real average for $\hat{\cal O}$ at any time $t$ 
by means of the simple expression $\langle \hat{\cal O} \rangle_Q^{BA}$, 
though it is generically complex by definition. 
In addition, we emphasize that, in the case of non-Hermitian Hamiltonians, we have obtained 
a $Q$-Hermitian Hamiltonian.

In the usual theory, i.e., the future-not-included RAT, 
the expectation value of $\hat{\cal O}$, $\langle \hat{\cal O} \rangle^{AA}$, is constructed to be real 
for a Hermitian operator $\hat{\cal O}$ by definition. 
Similarly, even in the future-not-included CAT, 
$\langle \hat{\cal O} \rangle_Q^{AA}$ is real 
for a $Q$-Hermitian operator $\hat{\cal O}$. 
On the other hand, in the future-included CAT and RAT, 
$\langle \hat{\cal O} \rangle_Q^{BA}$ is not adjusted so, 
but it becomes real by our natural way of thinking, the maximization principle. 
In addition, $\langle \hat{\cal O} \rangle_Q^{BA}$ is expressed 
more elegantly than $\langle \hat{\cal O} \rangle_Q^{AA}$ in the functional integral form: 
\begin{equation}
\langle \hat{\cal O} \rangle^{BA}_Q 
= \frac{\int {\cal D} \text{path}~ \psi_B^* \psi_A Q {\cal O} e^{ \frac{i}{\hbar} S[\text{path}] } }{\int {\cal D}  \text{path}~ \psi_B^* \psi_A Q e^{\frac{i}{\hbar} S[\text{path}]  }}. \label{funcint}
\end{equation}
In the future-not-included theories $\langle \hat{\cal O} \rangle_Q^{AA}$ does not have 
such a full functional integral expression for all time. 
Therefore, $\langle \hat{\cal O} \rangle_Q^{BA}$ seems to be more natural than 
$\langle \hat{\cal O} \rangle_Q^{AA}$, and we can speculate that the fundamental physics is given by 
$\langle \hat{\cal O} \rangle_Q^{BA}$ in the future-included theories 
rather than by $\langle \hat{\cal O} \rangle_Q^{AA}$ in the future-not-included theories. 
This interpretation provides a more direct connection of functional integrals to measurable physics.

In such future-included theories we are naturally motivated to consider the maximization principle. 
If we do not use it, $\langle \hat{\cal O} \rangle_Q^{BA}$, 
which is expected to have a role of an expectation value in the future-included theories, 
is generically complex by definition not only in the CAT but also in the RAT. 
This situation is analogous to the usual classical physics, where 
classical solutions are generically complex, unless we impose an initial condition giving the reality. 
Therefore, the maximization principle could be regarded as a special type of initial (or final) condition. 
Indeed, in the case of the future-included CAT, it specifies   
a unique class of combinations of $| A(T_A) \rangle$ and $| B(T_B) \rangle$. 
On the other hand, in the case of the future-included RAT, 
the maximization principle does not specify such a unique class, 
but only gives the proportionality relation: Eq.(\ref{BmaxphaseA}), and thus  
leaves the initial condition to be chosen arbitrarily. 
This is in contrast to the case of the future-included CAT. 
Thus the specification of the future and past states by the maximization principle 
is more ambiguous in the RAT than in the CAT. 
In this sense, the future-included CAT seems to be nicer than the future-included RAT, 
though it still requires a bit of phenomenological adjustment of 
the imaginary part of the action to 
get a cosmologically or experimentally good initial condition, 
and also suggests a periodic universe. 

Therefore, we speculate that the functional integral formalism of quantum theory 
would be most elegant in the future-included CAT. 
Though the future-included CAT looks very exotic, it cannot be excluded 
from a phenomenological point of view\cite{Nagao:2012mj, Nagao:2012ye}. 
Only the maximization principle would be needed in addition to the imaginary part of the action. 
The future-included CAT supplemented with the maximization principle 
could provide a unification of an 
initial condition prediction and an equation of motion.

\section*{Acknowledgements}
K.N. would like to thank the members and visitors of NBI for their hospitality 
during his visits to Copenhagen, and the organizers of the workshop Bled 2017 
for giving him the opportunity to contribute to the proceedings of the workshop.   
H.B.N. is thankful to NBI for allowing him to work at the institute as emeritus, 
and Matja\v{z} Breskvar for economic support to visit the Bled Conference.



\begin{thebibliography}{99}


\bibitem{Bled2006}
H.~B.~Nielsen and M.~Ninomiya, 
Proc. Bled 2006: What Comes Beyond the Standard Models, 
pp.87-124 (2006) 
[arXiv:hep-ph/0612250]. 




\bibitem{Nielsen:2007ak}
  H.~B.~Nielsen and M.~Ninomiya,
  Int.\ J.\ Mod.\ Phys.\  A {\bf 23}, 919 (2008). 
%


\bibitem{Nielsen:2008cm}
  H.~B.~Nielsen and M.~Ninomiya, 
  Int.\ J.\ Mod.\ Phys.\  A {\bf 24}, 3945 (2009).
%







\bibitem{Nielsen:2005ub}
  H.~B.~Nielsen and M.~Ninomiya,
  Prog.\ Theor.\ Phys.\  {\bf 116}, 851 (2007). 



\bibitem{Nielsen:2007mj}
H.~B.~Nielsen and M.~Ninomiya, 
Proc. Bled 2007: What Comes Beyond the Standard Models, pp.144-185 (2007) [arXiv:0711.3080 [hep-ph]].



\bibitem{newer1}
  H.~B.~Nielsen and M.~Ninomiya,
  arXiv:0910.0359 [hep-ph]. 



\bibitem{Vaxjo2009}
H.~B.~Nielsen, 
Found. Phys. {\bf 41}, 608 (2011) [arXiv:0911.4005[quant-ph]]. 






\bibitem{newer2}
H.~B.~Nielsen and M.~Ninomiya, 
Proc. Bled 2010: What Comes Beyond the Standard Models, 
pp.138-157 (2010) [arXiv:1008.0464 [physics.gen-ph]]. 

  





\bibitem{Nielsen2010qq}
H.~B.~Nielsen,
arXiv:1006.2455 [physic.gen-ph].


\bibitem{degenerate}
H.~B.~Nielsen and M.~Ninomiya,
arXiv:hep-th/0701018.





\bibitem{Nielsen2009hq}
  H.~B.~Nielsen,
arXiv:0911.3859 [gr-qc].




\bibitem{Bled2010B}
H.~B.~Nielsen, M.~S.~Mankoc~Borstnik, K.~Nagao, and G.~Moultaka, 
%
Proc. Bled 2010: What Comes Beyond the Standard Models, 
pp.211-216 (2010) [arXiv:1012.0224 [hep-ph]]. 














\bibitem{Nagao:2010xu}
K.~Nagao and H.~B.~Nielsen,
Prog.\ Theor.\ Phys. {\bf 125}, 633 (2011).




\bibitem{Nagao:2011za}
  K.~Nagao and H.~B.~Nielsen,
Prog.\ Theor.\ Phys. {\bf 126}, 1021 (2011); 
{\bf 127}, 1131 (2012) [erratum]. 





\bibitem{Nagao:2011is}
  K.~Nagao and H.~B.~Nielsen, 
   Int.\ J.\ Mod.\ Phys.\ A{\bf 27}, 1250076 (2012). 






\bibitem{Bender:1998ke}
  C.~M.~Bender and S.~Boettcher,
  Phys.\ Rev.\ Lett.\  {\bf 80}, 5243 (1998).

\bibitem{Bender:1998gh}
  C.~M.~Bender, S.~Boettcher, and P.~Meisinger,
  J.\ Math.\ Phys.\  {\bf 40}, 2201 (1999).


\bibitem{Bender:2011ke} 
  C.~M.~Bender and P.~D.~Mannheim,
  Phys.\ Rev.\ D {\bf 84}, 105038 (2011).







\bibitem{Mostafazadeh_CPT_ip_2002}
A. Mostafazadeh, 
J.\ Math.\ Phys.\ {\bf 43}, 3944 (2002).  
%



\bibitem{Mostafazadeh_CPT_ip_2003}
A. Mostafazadeh, 
J.\ Math.\ Phys.\ {\bf 44}, 974 (2003).




\bibitem{Nagao:2013eda} 
  K.~Nagao and H.~B.~Nielsen,
Prog.\ Theor.\ Exp.\ Phys. {\bf 2013}, 073A03 (2013). 






\bibitem{Nagao:2017ecx} 
  K.~Nagao and H.~B.~Nielsen,
  arXiv:1709.10179 [quant-ph].







\bibitem{Nagao:2012mj} 
  K.~Nagao and H.~B.~Nielsen,
  Prog.\ Theor.\ Exp.\ Phys. {\bf 2013}, 023B04 (2013). 


\bibitem{Nagao:2012ye} 
  K.~Nagao and H.~B.~Nielsen, 
Proc. Bled 2012: What Comes Beyond the Standard Models, pp.86-93 (2012) [arXiv:1211.7269 [quant-ph]].  






		
\bibitem{AAV}
	Y. Aharonov, D. Z. Albert, and L. Vaidman,
	Phys. Rev. Lett. 
	{\bf 60}, 1351 (1988).





\bibitem{review_wv}
	Y. Aharonov, S. Popescu, and J. Tollaksen,
	Phys. Today 
	{\bf 63}, 27 (2010). 









\bibitem{Nagao:2015bya} 
  K.~Nagao and H.~B.~Nielsen,
Prog.\ Theor.\ Exp.\ Phys. {\bf 2015}, 051B01 (2015).

 






\bibitem{Nagao:2017cpl} 
  K.~Nagao and H.~B.~Nielsen,
Prog.\ Theor.\ Exp.\ Phys. {\bf 2017}, 081B01 (2017).

\bibitem{Nagao:2017book} 
  K.~Nagao and H.~B.~Nielsen,
``Fundamentals of Quantum Complex Action Theory'', 
to be published by Lambert Academic Publishing. 





\bibitem{Geyer}
F. G. Scholtz, H. B. Geyer, and F. J. W. Hahne, Ann. Phys. {\bf 213}, 74 (1992). 




\end{thebibliography}
\end{document}